# Astrophotonics: molding the flow of light in astronomical instruments

**JOSS BLAND-HAWTHORN**[1,2,3] **AND SERGIO G. LEON-SAVAL**[1,2,3]

[1] *Sydney Astrophotonic Instrumentation Labs, School of Physics, University of Sydney, NSW, Australia*
[2] *Institute of Photonics and Optical Science, School of Physics, University of Sydney, NSW, Australia*
[3] *Sydney Institute for Astronomy, School of Physics, University of Sydney, NSW, Australia*
*\*jbh@physics.usyd.edu.au*

**Abstract:** Since its emergence two decades ago, astrophotonics has found broad application in scientific instruments at many institutions worldwide. The case for astrophotonics becomes more compelling as telescopes push for AO-assisted, diffraction-limited performance, a mode of observing that is central to the next-generation of extremely large telescopes (ELTs). Even AO systems are beginning to incorporate advanced photonic principles as the community pushes for higher performance and more complex guide-star configurations. Photonic instruments like Gravity on the Very Large Telescope achieve milliarcsec resolution at 2000 nm which would be very difficult to achieve with conventional optics. While space photonics is not reviewed here, we foresee that remote sensing platforms will become a major beneficiary of astrophotonic components in the years ahead. The field has 'given back' with the development of new technologies (e.g. photonic lantern, large area multi-core fibres) already finding widespread use in other fields; *Google Scholar* lists more than 400 research papers making reference to this technology. This short review covers representative key developments since the 2009 *Focus* issue on Astrophotonics.

## References and links

## 1. Introduction

Photonics provides a playground in which light can be manipulated in ways that go far beyond what is possible in optics. In some instances, there are analogies. For example, broadly speaking, the array waveguide grating (AWG) simulates the action of a collimating lens feeding light to a grating that diffracts the signal towards an output camera lens, and it does this efficiently in a compact planar geometry. But there are hidden advantages that are difficult to match with conventional optics and gratings. The AWG can be modified to achieve very different orders of interference $m$ by a change in the track spacing within the same compact space envelope and geometry [1]. Other examples have no easy analogy in optics. The single-mode fibre (SMF) is a perfect spatial filter when illuminated with a gaussian beam accepting only the component of the wavefront that is orthogonal to the direction of the fibre axis [2]. This is why SMFs are favoured in stellar interferometry and exoplanetary detection. The photonic lantern [3] has absolutely no counterpart in optics.

Here we provide a far from exhaustive, cursory review of developments since the 2009 *Focus* issue to illustrate the wide range of activities taking place. This format limits the length of reference lists but a search under *Google Scholar* can identify links to all of the topics under discussion, hence our liberal use of acronyms and proper nouns.

## 2. Coupling and cleaning light

When a plane wave with wavelength $\lambda$ is focused onto a telescope focal plane, the transmitted light evolves along its path to project to a series of concentric rings with a central intensity peak. This Airy pattern emerges if the telescope is diffraction-limited at the wavelength $\lambda$. If the FWHM of the inner peak is approximately matched to the modal diameter of an SMF, the coupling can be no better than about 78% [4]. But the modal properties of the refractive index profile can be adapted to couple up to 95% of the incoming light [5], excluding Fresnel losses due to reflection at the face. Interestingly, the Airy-like mode is not the fundamental mode of the fibre, but a higher order longitudinal mode with a guidance mechanism having strong similarities to photonic bandgap guidance [5].

The coupling is more challenging when fibres are placed at the focus of a front-line telescope. Problems include image motion, decentering and apertures with central obstructions typical of most telescopes (e.g. Cassegrain or Nasmyth optics). The last two issues are discussed here [6]; almost complete coupling can be achieved in both cases with few mode fibres. Guyon [7] goes one step further by showing that a pair of phase-induced apodized aperture (PIAA) lenses can fill in the central hole in the beam due to the obstruction to produce a Gaussian output beam. The extreme AO system (SCExAO) utilizing PIAA lenses and a 4000-actuator deformable mirror at the Subaru Telescope in Hawaii has achieved better than 60% coupling onto a single-mode fibre at near IR wavelengths [8]. This is a remarkable achievement, i.e. *the reflected light of an 8m diameter telescope is mostly coupled to the modal diameter (10 microns) of a single-mode fibre*.

The astrophotonics community recognizes the problem of coupling a 2D waveguide to an optical fibre to be of paramount importance in the quest to build high-performance photonic instruments [8-10]. Industry uses both grating couplers and tapered couplers to achieve good performance where the latter has a better broadband response. Recently, the University of Maryland engineering and astronomy groups working together have achieved 93% butt-coupling efficienctly [11] from an SMF to an SM silicon nitride ($Si_3N_4$) waveguide using a $SiO_2$ on Si platform. The taper couplers are made with e-beam deposition to achieve the 100 nm features. This is an area where nanoscale engineering and nanophotonics in general will improve the statecraft in future years.

One of the real challenges involved in precision spectroscopy (e.g. HARPS at the ESO 3.6m telescope) is ensuring a perfectly stable output Gaussian beam from a fibre in the presence of image motion at the input face. Even without image motion, fibres produce a wavelength-dependent speckle (granular) pattern resulting from interference of the many propagating modes with different relative phases. This is not a problem with SMFs which allow only one fundamental mode of propagation, a key reason why these are now favoured as the input to high-precision radial velocity spectrographs. For a few-mode or multimode fibre, this presents a real challenge, particularly for radial modes. Multimode fibres scramble in azimuth fairly efficiently, but this is not true for the radial modes. Spreading the power evenly across guided modes in radius is very challenging, particularly if the goal is to conserve the beam properties of the fibre and not to lose light through focal ratio degradation (NA upscattering). Stürmer et al [12] reviews different fibre geometries that achieve mode scrambling, e.g. fibres with D-shaped cross sections, inspired by chaos theory, outperform polygonal fibres and behave rather well.

## 3. Transporting light

Since the 1970s, astronomers have observed multiple sources simultaneously by aligning multi-mode fibres with celestial targets imaged onto the telescope focal plane. This has led to decade-long surveys of stars and galaxies with wide-field instruments (SDSS, 2dF) involving millions of sources observed over the full sky. These huge data sets have produced some of the highest cited research papers in astrophysics.

Several major instruments on front-line telescopes receive light at the face of fibre bundles rather than individual fibres. Sometimes, microlens or macrolens arrays are called for to match the input beam to the natural numerical aperture (NA) of the fibre bundle. One of these technologies, the hexabundle [13], makes use of photonic principles (lightly fused, reduced cladding) to maximize the covering fraction (75-85%) of the stacked fibres.

Polymethyl methacrylate (PMMA) and silica coherent fibre bundles – widely used in biomedical endoscopy – have been explored for accurate field acquisition and guiding, wavefront sensing, narrow-band imaging, aperture masking, and speckle imaging [14]. Astronomical bundles typically require the material to be flexible due to mechanical motion between the telescope and instrument which provides the designer with unique challenges. Other photonic light transport applications include large aperture, "endlessly single-mode" fibres to carry high laser power in AO systems needed for generating artificial stars in the atmospheric sodium layer at 90 km altitude [15]. A new technology being driven now by astrophotonics is large area multicore fibres. Once light is coupled into the cores (e.g. using a photonic lantern), it can travel great distances without degradation.

## 4. Combining light

Labadie [16] has provided a comprehensive update on progress with optical beam combining in astrophotonics. For completeness, we recap a few developments but stress that this is a burgeoning field in astrophotonics which could easily fill its own *Focus* issue. Complex waveguides with highly sophisticated metrology (e.g. path compensation) are now routinely used in interferometric instruments like Gravity and PIONIER on the Very Large Telescope

(VLTI) in Chile, Dragonfly and the GLINT photonic nuller on the Subaru Telescope in Hawaii. Early instruments in the 1990s like FLUOR used X-couplers and Y-couplers to gather both the interferometric and photometric signal of the target source. Later instruments like IONIC used integrated waveguide optics to combine the beams from two or more telescopes. This instrument achieved the first closure-phase results from observations of spectroscopic binary stars. Their use of $SiO_2$-Si waveguide technology ensured an end-to-end insertion loss of only 30% at 2000 nm.

Quite apart from the 2015 direct detection of gravitational waves and merging black holes, the Laser Interferometric Gravitational-Wave Observatory (LIGO) sets a truly remarkable new benchmark for laser metrology. In a nutshell, LIGO is good news for both AO systems and astrophotonics. The metrology of Gravity is arguably the best of a telescope-based instrument [17] which is demanded by their requirements on narrow-angle astrometry through phase-shifting interferometry to detect stellar motions near the Galactic Centre supermassive black hole (Sgr A*).

There is an ongoing effort to extend the integrated optics approach in beam combining to mid-infrared wavelengths [18]. This includes consideration of new platforms (Si on insulator, Si on sapphire) and photonic structures (e.g. Kagome lattices, large area hollow air-core fibres). Many new spin-off technologies like microheaters for path compensation have broad applications in other areas of astrophotonics. Moreover, the push to shorter wavelengths is also under way [19].

## 5. Dividing light

The photonic lantern [3] is a specialty optical transition device that is multimoded (single aperture) at one end and evolves along its length through at least one adiabatic taper to a different structure, such as an array of multiple SMF or few-moded outputs, or one or several MCF outputs. In the original photonic lantern design (i.e. multimode/single-mode/multimode), this final structure then underwent a reverse adiabatic taper back to a multimode output. This was to permit single-mode action in a multimode fibre [20], e.g. by printing an identical Bragg grating across all multiple single-mode waveguides in the bridge region. In general, light can be transformed, divided and/or functionalised by a conventional single-mode photonic device (e.g. interferometers, ring resonators, optical couplers, etc.), be collected up and fed back into a photonic lantern to generate multimoded input and output; alternatively, the light on its SMF, few-moded, or MVF form can be directed to a low-noise detector or any other type of sensor for analysis. Temporal, spatial, and wavelength dependent functions can now be exploited in astronomy to pursue the desired science.

Photonic lantern technology has proved to be versatile. For example, while the original photonic lantern design used identical single-mode waveguides in the bridge region, they need not be. Asymmetric lanterns, also known as mode-selective structures, arose to meet the needs of new optical transmission architectures in the telecomm industry [21]. Contrary to bulk optic approaches for dividing and transforming light, photonic lanterns use modal properties and waveguide optic principles to work. Thus, many different optical transitions can be achieved by the photonic lantern technology as long as the entropy of the system, i.e. number of modes is conserved. They can be used as "light dividers" to allow for broadband low-loss interface between multimode, few-mode and single-mode systems, thus allowing a waveguide transition from one to other as required.

Nowadays, this same technology can be used to convert a massively multimoded (single aperture) input to many fewer-moded multimode outputs. For example, this type of lantern can be employed as a very low-loss optical fibre slicer, although more work is needed to understand the optical properties as a function of wavelength. Multimode to multimode photonic lanterns not only provide a practical way to use large-core fibres and hexabundles in large focal plane telescopes, but also a way forward for highly-multimoded photonic technologies in next generation astronomical instrumentation. Future work, and current areas of research, on photonic lanterns as light dividers in astronomy will include: low modal noise; focal ratio degradation free input/output optical fibre systems; and all-fibre NA converters that conserve étendue.

## 6. Dispersing light

Conventional optical approaches to dispersing light are well documented. As telescopes get larger, the dispersing instruments grow at the same rate if the instrument performance is tied to natural atmospheric (seeing) conditions. But in the current era, most leading and future telescopes are investing heavily in adaptive optics with a view to achieving (near) diffraction limited performance, particularly at infrared wavelengths. This provides a crucial opening for astrophotonics since a diffraction-limited beam can, in principle, be coupled efficiently to a single-mode waveguide. This has led to much interest in photonic methods for dispersing light and reduced volume microspectrographs [22]. Some of the most exciting developments involve the use of array waveguide gratings originally developed for DWMD in telecom [23, 24].

The different approaches to astronomical spectroscopy can be captured succinctly [25] – the power spectral density of a stationary random process is the Fourier transform (FT) of the corresponding autocorrelation function (Wiener-Kitchin theorem). Photonics allows for a myriad of new ways to analyze the autocorrelation of light propagation in time, some of which have no analogy in optics, but others are integrated optic versions of bulk optical systems (e.g. FT spectrometer). Like interferometry, this is a huge topic easily deserving its own *Focus* issue. Some interesting developments are semi-integrated double pass dispersion, side holographic dispersion, photonic crystal superprisms, zigzag spectroscopy, MEMS-based spectroscopy, Fresnel microspectrometry, plasmonic photon sorting, stationary wave integrated FTS, leaky loop integrated FTS, all-in fibre wavemeter FTS, arrayed Mach-Zehnder interferometry, volume Bragg gratings and a dazzling armada of other variants [25]. All things become possible when one considers the interaction of light with materials.

## 7. Filtering light

Photonics provides us with an extraordinarily powerful medium for constructing complex optical filters, e.g. to suppress hundreds of non-periodically spaced, unwanted frequencies. An example from astronomy is the long-standing problem of suppressing the bright forest of atmospheric emission lines arising from the Earth's atmosphere. Complex photonic filters at microwave frequencies have long existed and are the staple of modern radio noise mitigation systems. At optical and infrared wavelengths, fibre Bragg gratings (FBGs) printed into SMF tracks are the only viable technology at the present time, particularly if you want to achieve low overall insertion loss while maintaining a high degree of suppression in the noise spikes.

The principle of the FBG is that light propagating along an optical fibre can be made to undergo Fresnel reflections at many refractive index increments printed onto the fibre core. If the modulation describes a grating, light can be made to reflect back along the full length of the grating. The grating is defined by a complex phase and modulation amplitude along its length. The complex filter observed in transmission arises from interference between the forward-propagating field and the backward-propagating field across the grating. For these reflections to add up coherently, the fibre must be single-moded (i.e. propagation vector aligned with the fibre axis). Thus the photonic lantern was introduced to allow for single-mode action into a multimode fibre.

The most advanced work in this area comes from our group [20] working with lab-scale Mach-Zehnder and Michelson interferometers to achieve highly structured refractive index modulations along a SMF track [26]. This work led to 105 notches in the window 1450-1700 nm with better than 30 dB peak suppression. Our latest foray is to print these complex gratings into multi-core fibres in order to minimize the effort of making these OH-suppressing fibres. The work is in its infancy but good progress is being made [27].

With reference to the last section, one can also consider the prospect of complex filtering integrated with a dispersing element. In recent years, there has been some work in this direction. Spelaniak and colleagues [28] have achieve a 3D integrated photonic chip using a solid photonic lantern and etched Bragg gratings made using a femtosecond direct-write technique in boro-aluminosilicate Eagle 2000 glass. Only a few suppression dips were possible with a peak suppression of 5 dB, but this approach shows promise. The Maryland group has now managed to print 47 notches over 1530-60 nm window with peak suppression of about 15 dB, more than doubling the number in [11]. The technology is based on a Bragg waveguide grating achieved by depositing $Si_3N_4$ onto a Si wafer: each track has a complex ridge structure in order to achieve the non-periodic interference. This is much closer to the requirements of OH suppression but the overall insertion loss is much higher than FBG technology at the present time.

## 8. Calibrating light

Photonic combs came to attention in astronomy a decade ago in the context of improving on the iodine cell method for precision radial velocities to assist with the search for exoplanets. Here the imprint of the iodine absorption spectrum, with its accurately calibrated atomic, spectrum overlays the stellar light. The first realizations of laser combs were unwieldy, unstable and expensive - big powerful lasers were fed into large etalon cavities in order to remove tines from the densely bunched optical frequency comb [29]. Compact ring resonators [30] and fibre etalons [31] have also been discussed as possible calibrators for astronomy although scientific results have not been presented to date.

We refer to photonic combs collectively as all devices that seek to produce a periodic output in frequency for the purposes of instrument calibration, both in terms of wavelength calibration and PSF mapping. This includes laser combs, fibre etalon combs, ring resonator combs and so on. There are many ways to achieve this sort of output as described in recent reviews [32]. We reserve the term "frequency comb" for applications that require the highest precision and therefore lock to microwave atomic clocks through an electronic feedback loop. This technique was brilliantly exploited in atomic physics and led to the 2005 Nobel Prize being awarded to Hänsch and Hall.

A demonstration of the power of the photonic comb in removing the systematic aberrations of a wide-field spectrograph is given in [31]. We discuss the precision locking of fibre etalons in a companion paper [33]. By analogy with the iodine cell, there is a case for fibre etalon which produces narrow notches in absorption, particularly if the notches achieve >30 dB suppression. Such a response can be used to define the true spectrophotometric baseline of the instrument in the presence of scattered light. There is also a demand for an extremely stable spectrophotometric standard whose shape can be "tuned" to the celestial source under study. Both constitute unsolved problems at the present time which we anticipate can be resolved using photonic mechanisms.

## 9. Converting light

Radio astronomy has a long history of heterodyne techniques where closely spaced high frequency signals are combined to produce a lower beat frequency where electronic processing is easier and less expensive to achieve. Not to be outdone, astrophotonics has begun to exploit optical frequency *up*-conversion to allow near-IR signals to be transferred to the visible where high-performance CCD detection and processing is far less expensive. Frequency up-conversion is a widely used technique based on sum frequency generation in a non-linear optical medium in which signal light from one wavelength is converted to another wavelength. A team at the University of Limoges has already obtained on-sky results in the H-band, in addition to a lab demonstration at $\lambda 3.4\mu m$ converting into the near-IR [34, 35]. Their conversion stage uses a lithium niobate, tapered waveguide which undergoes "periodic poling" with a pulsed UV laser; this enforces the quasi-phase-matching condition by enhancing the 3$^{rd}$ order non-linearity. At present, the overall efficiency is not high but, at the spectral resolution delivered, the overall sensitivity (≤1%) could be competitive with Gravity and MATISSE working with the VLTI [36].

## 10. Discussion

We have left many other topics undone, e.g. developments that exploit higher moments of light propagation including linear and circular polarization. There are novel advances in the use of photon orbital angular momentum (OAM) which exploits the field spatial distribution of the optical beam as distinct from the polarization components. One of the more spectacular developments is vortex coronagraphy. This technique suppresses a bright star by inducing a helical phase front in the optical beam such that the beam axis has a phase singularity, thus a "hole" in the propagating wave front. Orbiting stars, planets and dust clouds have now been found around nearby stars using this method [37].

There is some controversy in the literature about the usefulness of the OAM signal over astronomical distances. But if you consider the photon to be one unit of excitation of a specific mode of the electromagnetic field, there is no confusion – even *single photons* carry OAM information. OAM signatures have been claimed in stellar measurements [38] but the role of the atmosphere needs to be considered. We suggest that OAM measurement may be key to future AO systems as an independent probe of their performance. There are also interesting developments in so-called "quantum astronomy" where higher order moments of light are exploited in photonic Hanbury Brown Twiss (HBT) experiments, for example. We recall the controversy that surrounded the usefulness of HBT in the early years [39], now the fundamental underpinning of quantum optics. Higher order moments are easily measured with photonic technologies and thus we suspect there is scope for real advances here in astrophotonics.

Astrophotonics to date has assumed that the light is intrinsically unpolarized but this is not a limitation in principle. Photonic devices can easily sustain the integrity of the TE and TM modes, at least in principle. More broadly, a fast way into the field for a young scientist is to consider the polarizing properties of any photonic device. Too little work has been done in this area. We often assume our devices are blind to polarization but they rarely are, e.g. light propagation in multi-track waveguides [40]. Moreover, optical switching offers interesting possibilities in future instruments [41].

We proffer a note of caution. Adaptive optics, astronomical lasers and photonic combs (e.g. frequency combs, ring resonators) are now widely used across astronomy. These advances make heavy use of developments within the photonics and optics industry over fifty years. But in all cases, substantial changes and developments were necessary for the technology to be useful. For example, large-format micromirror devices or microchannel plates available today have all arisen within astronomy. The same of course is true for the highest performance wide-area detectors sustained for decades by large grants to the astronomy community working closely with industry (e.g. e2V). As discussed in past reviews, the ideal detectors for astrophotonics (e.g. curved detector plane, linear array format with ~10$^4$ pixels with sizes of order a few microns) are not generally available, but there has been good progress on high performance, curved silicon CMOS arrays at Microsoft Corp.

There are many key components (e.g. spatial light modulators), beloved of the photonics community, whose optical performance is not good enough for more widespread use in astronomical instruments. These can be improved with significant investment. In time, as the case becomes more compelling, we can expect to see more key technologies (e.g. optical circulators) being adopted by the astronomical and space research communities.

A major success of astrophotonics and space photonics is the demonstrated efficacy in keeping instruments as compact as possible [22]. Some of the field's newest technologies feature in the University of Sydney's photonic cubesat that launched from Cape Canaveral in April 2017, then deployed from the International Space Station in May 2017. This is also one of the goals of the iCubeSat programme at Paris Observatory with their upcoming proposal to measure transit photometry towards $\beta$ Pictoris using single-mode fibres.

We foresee that compact photonic instruments (e.g. microspectrographs) will be able to exploit subwavelength photonics, currently in vogue, made possible with engineered nanostructured materials. These include: (i) photonic crystal fibres (PCFs) where an index modulation is introduced by a crystal structure in the cross-sectional plane; (ii) subwavelength-grating waveguides (SWG) which are periodic metallic or dielectric waveguides; (iii) metamaterials which are nanostructured materials that support a designer's choice of permittivity and permeability values, inducing artificial magnetic fields, thereby yielding a negative refractive index, chiral or bi-anisotropic propagation, and so forth. This will allow for better on-chip optical interconnects, improved binary, diffractive and holographic gratings, and better off-axis performance in diffractive optical elements.

In our view, a future that is dominated by astrophotonics, rather than bulk optics, is inevitable. In an era when all telescopes work at or near the diffraction limit, in space or on the ground, there is no advantage and only risk associated with large components. This assumes, of course, by analogy with the other commercial technologies, that industry continues to drive the cost of photonic components downwards. An era of fully integrated photonic instruments is only a matter of time.

## Acknowledgments

This article has benefitted from two excellent referees to whom we owe our thanks. We continue to be inspired by the many scientists engaged in astrophotonics and thank them for the support and motivation to write this review. The restricted format has not allowed us to cover the full breadth of extraordinary developments worldwide. We are indebted to Miles Padgett and Michael Berry for their insights.

## Funding

The SAIL labs are supported by JBH's Laureate Fellowship FL140100278 from the Australian Research Council and from DVC-R Strategic Funding awarded by the University of Sydney.